\documentclass[prl,twocolumn,showpacs]{revtex4}
\usepackage{amsmath, amsgen,amstext,amsbsy,amsopn,amsthm, amssymb}
\usepackage[dvips]{graphicx}

\theoremstyle{definition}

\usepackage{color}

\def\*{{\phantom *}}
\begin{document}
\title{The second critical point for the Perfect Bose gas in quasi-one-dimensional traps}

\author{Mathieu Beau}
\author{Valentin A. Zagrebnov}
\affiliation{Universit\'e de la M\'editerran\'ee and Centre de Physique
Th\'eorique - UMR 6207
\\ Luminy - Case 907,
13288 Marseille, Cedex 09, France}

\date{February 18, 2010}

\begin{abstract}
We present a new model of quasi-one-dimensional trap with some unknown physical predictions about a second
transition, including about a change in fractions of condensed coherence lengths due to the existence of a second critical temperature $T_m<T_c$.
If this physical model is acceptable, we want to challenge experimental physicists in this regard.
\end{abstract}

\pacs{05.30.Jp, 03.75.Hh}

\maketitle
\noindent 1. It is well known since 1925 \cite{Einstein} for an ideal gas of identical bosons of mass $m$ in a three dimensional
cubic box $\Lambda = L^3$  described
the grand canonical ensemble $ (V, T, \mu) $ with $ T $ the temperature and $ \mu $ the chemical potential,
there exists a critical density of thermal gas saturation of $ \rho_c = \zeta (3 / 2) / \lambda_\beta ^ 3 $
where $ \lambda_\beta = \hbar \sqrt{2 \pi \beta / m} $ is the thermal length and $ \beta = 1/k_BT $ .
The hypothesis of A. Einstein back in 1938 \cite{London} by F. London assumed that for a particle density $ \rho $
above the critical density $ \rho_c $ surplux the particles $ \rho-\rho_c $  starts in the ground state mode $ k = 0$ .
Thus we note that $ \rho_{k = 0} = \rho-\rho_c $ for $ \rho> \rho_c $.
But this argument macroscopic population of the ground state depends drastically on the geometry of the box.
This is only first discovered H. Casimir \cite{Casimir} in 1968 for a certain geometry anisotropic
and what has motivated the work of M. van den Berg, J. Lewis and J. Pule from 1981 \cite{vdBL} - \cite{vdBLL}.
These are generalized the concept of Bose-Einstein condensate defining density in thermodynamic limit
for an ideal Bose gas in a rectangular box any $ \Lambda = L_x \times L_y \times L_z$:
\begin{eqnarray}
\rho_0=\lim_{\eta\rightarrow0}\lim_{L_x,L_y,L_z\rightarrow\infty}\sum_{\|k\|\leqslant\eta}\rho_k=\rho-\rho_c.
\end{eqnarray}

This definition includes all possibilities from a condensate, given here from \cite{vdBLP}
and revisited \cite{Beau} in reference to new terminologies \cite{MHL}:

(i) The condensate conventional or usual \cite{Beau}, \cite{MHL} formed by the ground state (type I \cite{vdBLP}).

(ii) The fragmented condensate \cite{Beau}, \cite{MHL} if the generalized condensate is distributed over a set (finite / infinite) modes macroscopically occupied
in an energy band near the fundamental mode ie $ N_0 = \sum_{k \leqslant k_c} N_k $, with $ N_k = O (N) $ (Type I / II \cite{vdBLP}).

(iii) The quasi-condensate \cite{Beau}, \cite{MHL} if the condensate is distributed over a generalized set of modes occupied mesoscopic
in an energy band near the fundamental mode ie $ N_0 = \sum_{k \leqslant k_c} N_k $, with $N_k = O (N ^ \delta),\ \delta <1 $ (type III \cite{vdBLP}).

In 1983 \cite{vdB}, van den Berg proposed a box model three-dimensional anisotropic exponentially in both directions
to define a second critical density $ \rho_m $  separating a regime of condensate generalized type III (for $ \rho_c <\rho <\rho_m$ )
towards a regime of generalized type I condensate (for $ \rho> \rho_m$ ).
Lately, we have reexamined this model \cite{BeauZ} questions that we call two-dimensional and we have shown that this
a transition between two regimes: quasi-condensate and coexistence between a regime of conventional condensate and quasi-condensate,
notably by showing that the second critical density $ \rho_m $  corresponds to a saturation density of gas near condensation
similarly to the saturation of thermal gas for $ \rho_c $ .
In addition we have calculated the second critical temperature and determined the fractions similar condensed changed.
Then we calculated the effects of this transition on the coherence length for these two regimes of condensate generalized.

However this transition in the model boxes van den Berg is not the case for almost one-dimensional (quasi-1D).
Our finding presented here is the existence of this transition between two regimes for the ideal gas of bosons
in a harmonic trap quasi-1D.
The purpose of our letter is to show where is the transition between a quasi-BEC and BEC conventional
find the geometric model for the trap to his observation and to characterize this transition
(Fractions condensed, coherence lengths).
This theoretical prediction is interesting because we have seen in the experimental \cite{Gerbier}
and theoretical articles \cite{Petrov1}, \cite{Petrov2} the existence of a second critical temperature called $ T_\phi $
for quasi-1D environments, but calculated and explained differently from ours.

\noindent 2. Consider an ideal gas of bosons in a harmonic trap characterized by three-dimensional pulsation $ \omega_x, \omega_y, \omega_z$ :
where external potential is given by:
\begin{eqnarray}
V_{ext}(r)= \frac{1}{2} m\omega^{2}_{x} x^2 + \frac{1}{2} m\omega^{2}_{y} y^2 + \frac{1}{2} m\omega^{2}_{z} z^2 .
\end{eqnarray}

The Hamiltonian for a particle is given by \cite{Dalfovo}:
\begin{equation}\label{1D-harm-trap}
T_{\Lambda}^{(N=1)}= -\frac{\hbar^2}{2m}\Delta +V_{ext}(r)\ .
\end{equation}

The energy levels are then given by:
\begin{eqnarray}
\epsilon_{s_x s_y s_z}=\hbar\omega_z(s_z +\frac{1}{2})+\hbar\omega_z(s_z +\frac{1}{2})+\hbar\omega_z(s_z +\frac{1}{2}).
\end{eqnarray}

Let $ \mu $ the chemical potential of this system in the grand canonical ensemble,
its value must be strictly lower than the ground state to ensure the convergence
distribution the Bose-Einstein statistics showing the number average boson in each of the modes of expression
$ N_{s_x s_y s_z} = (\exp{\beta (\epsilon_{s_x s_y s_z} - \mu)} -1) ^{-1} $.
Thus we can define the effective chemical potential relative to ground level as $ \mu = \epsilon_{000} - \Delta \mu $.
When the potential $ \Delta \mu> 0 $ tends to 0, the thermal gas is saturated and we obtain a critical number of particles when
$ \omega_0 = (\omega_x \omega_y \omega_z)^{1/3} $ tends to 0:
\begin{eqnarray}\label{DR-n1-harm}
N&\simeq&\frac{1}{(\hbar\beta\omega_0)^3}\int_{\mathbb{R}^{3,+}}d^3\epsilon \frac{1}{e^{\beta(\epsilon+\Delta\mu)}-1}
{}\nonumber{}\\&<&\frac{\zeta(3)}{(\hbar\beta\omega_0)^3}=N_c.
\end{eqnarray}

The nature of the spectrum for this trap is different from that of a particle in a box,
that is why the type of geometry for the second transition will be different.
If we consider that this trap is anisotropic with a quasi-1D type "cigar" in the axis $ z $,
we must have the condition $ \omega_z \ll \omega_{\perp} $ with $ \omega_x = \omega_y = \omega_\perp $.
Usually \cite{Dalfovo}, \cite{Petrov1}, \cite{Petrov2} is taken as a model $ \omega_{\perp} = \lambda\omega_z$ with the condition $ \lambda \gg1$.
Thus we have an anisotropy level of modes in each direction for $ \hbar \omega_z \ll \hbar \omega_\perp $.
If in this case we have an effective chemical potential $ \hbar \omega_z \ll \Delta \mu \ll \hbar \omega_\perp $
then we can calculate the number of atoms in a condensate generalized (quasi-condensate) in an energy band near the fundamental on the axis $ z $:
\begin{eqnarray}\label{Eq3-harm}
N_{qbec}&\equiv&\sum_{s =(0,0,s_z)} {N_s}\simeq\frac{1}{\hbar\beta\omega_z}\int_\mathbb{R^{+}}d\epsilon_z\frac{1}{e^{\beta(\epsilon_z+\Delta\mu)}-1}
{}\nonumber\\{}&\simeq&- \frac{1}{\hbar\beta\omega_z} \ln [\beta \Delta\mu] =
{N} - N_c .
\end{eqnarray}

Thus for a macroscopic number of particles in the condensate, we must have
condition $ \Delta \mu = \beta^{-1} e^{- \hbar \beta \omega_z (N - N_c)} + \ldots $.
But there is another condition for completing the energetic part along the axis $ z $
because the chemical potential $ \Delta \mu $ determines the size of the energy band,
decreases when the number of particles $ N $ increases and therefore a second critical number of particles
noted $N_m$, the condition $ \hbar \omega_z \ll \Delta\mu $ is false.
For this model unlike gas trapped in a box, there is no global notion of "density" of particles as the containment
takes place throughout the three-dimensional space $ \mathbb{R}^3 $ .
However we can introduce a scaling for the number of particles $ n \equiv N \omega_0^3 $
which converges when $ N $ tends to infinity and when $ \omega_0 $ tends to 0 \cite{Dalfovo}.
We then find that when $ \hbar \omega \simeq \Delta \mu $ is a $ \beta \hbar \omega_z \simeq e^{- \hbar \beta (n_m-n_c) / \omega_\perp^2} $ .
To resolve this condition, we introduce the model of anisotropic quasi-1D trap following $ \omega_z = \omega_\perp e^{- \omega_c^2 / \omega_\perp^2} $
where $ \omega_c> 0 $ is a pulse of anisotropy exponentially.
Thus there is a limit to the number of particles in the condensate distributed over the energy axis $ z $:
\begin{eqnarray}\label{Nm}
N_m = N_c + \frac{\omega_c^2}{\hbar\beta\omega_0^3}.
\end{eqnarray}
Now if we go beyond this critical number $N_m$, we shall complete the ground state:
\begin{eqnarray}
N_{bec}\equiv N_{s=0}= \frac{1}{e^{\beta\Delta\mu}-1}\simeq\frac{1}{\beta\Delta \mu} = N-N_m,
\end{eqnarray}
and thus the chemical potential is $ \Delta \mu = 1 / (\beta (N-N_m)) $ and the value of the quasi-condensate is saturated:
\begin{eqnarray}
N_{Qbec}&\simeq& -\frac{1}{\hbar\beta\omega_z} \ln [\beta \Delta\mu] =N_m - N_c ,\nonumber
\end{eqnarray}
since $ N-N_c = (n-n_c) / \omega_0^3 $
and thus $ \omega_0^3 N_{Qbec} \simeq(\hbar \beta)^{-1} \omega_\perp^2 \ln[\omega_0^3 / (n-n_c)] \simeq \omega_c^2(\hbar \beta)^{-1}$

Note that this model of exponential anisotropy is necessary to limit "thermodynamics", ie when $ \omega_\perp $ tends to 0,
to define the second critical number of particles $ N_m $ which is the same order as the first critical number $ N_c $ (proportional to $ \omega_0^{-3} $).
For finite pulses in a given experimental situation, we must calculate the parameter $ \omega_c $
depending on the aspect ratio $ \lambda = e^{\omega_c^2 / \omega_\perp^2} $ and see if the ratio $ \omega_c / \omega_\perp $ is sufficiently
large to consider that one is in an exponential phase.
Usually if one considers a system where very anisotropic $ \lambda \gg1 $
to consider this point of view that the system is exponentially anisotropic, we have to have $ \ln (\lambda) \gg1$ . \\
\begin{figure}
\begin{center}
  \includegraphics[scale=0.3]{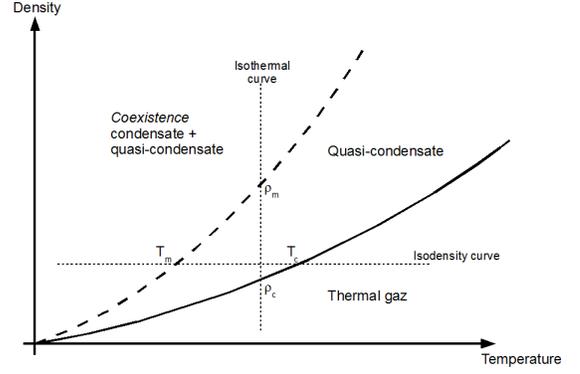}
  \caption{Illustration of phase diagram, see (\ref{DR-n1-harm}), (\ref{Nm})}
  \end{center}
\end{figure}
\noindent 3. In experiments with BEC, it is important to know the critical temperatures associated with
corresponding critical densities.
The \textit{first} critical temperatures: $T_c$, for a given particles number $N$is given by solutions of this equation:
$N = N_c=\zeta(3)/(\hbar\beta\omega_0)^3$.
Then we get:
\begin{equation}\label{Tc}
{T}_c=\frac{\hbar\omega_0}{k_B \zeta(3)^{1/3}}N^{1/3}\ .
\end{equation}
By the expression for the \textit{second} critical particles number one gets the following relation between the \textit{first} and the \textit{second}
critical temperatures because $T_m$ are given by solution of this equation $N=N_m=N_c+\omega_c^2/(\hbar\beta\omega_0^3)$:
\begin{eqnarray}\label{Tm}
{T}_{m}^{3}+{{\tau}}^2\ {T}_{m}&=&
{T}_{c}^{3} \ .
\end{eqnarray}
Here ${\tau} = \omega_c\hbar/k_B\zeta(3)^{1/2}$ is "effective" temperatures related to the corresponding geometrical shape.

\begin{figure}
\begin{center}
  \includegraphics[scale=0.2]{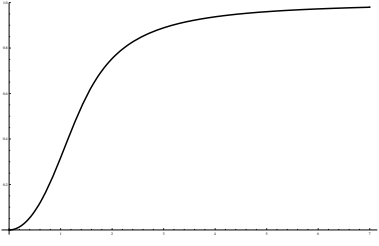}
 \caption{Curve illustrating (\ref{TmTm}) as a function of $\frac{\tau}{T_c}$.}\label{C3FigTmTc}
  \end{center}
\end{figure}

We can compute the relative difference between both critical temperature due to (\ref{Tm}) , see figure \ref{C3FigTmTc}:
\begin{eqnarray}\label{TmTm}
\frac{T_c - T_m}{T_c}=f(\frac{N_{\omega_c}}{N})
\end{eqnarray}
where $N_{\omega_c}=\frac{\omega_c^3\zeta(3)^{1/2}}{\omega_0^3}$ defined by $\frac{N_{\omega_c}}{N}=\frac{\tau^3}{T_c^3}$
and with:
\begin{eqnarray}
f(x)=1-\Omega(x)^{1/3}+\frac{x^{2/3}}{3}\Omega(x)^{-1/3}
\end{eqnarray}
where $\Omega(x)=\frac{1}{2}(1+\sqrt{1+\frac{4}{27}x^2})$.

\begin{figure}
\begin{center}
  \includegraphics[scale=0.2]{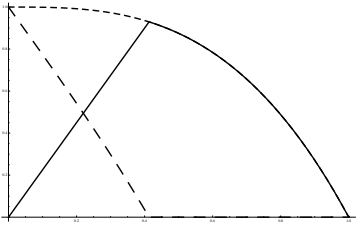}
  \caption{Illustration of condensate fractions (\ref{BEC-III-sg}) (---), (\ref{BEC-I-sg}) (- - -), (\ref{BEC-Tot-sg}) (-  -  -)
  with $\frac{T_c}{\tau}=\frac{2}{3}$}\label{CondensateFract}
 \end{center}
\end{figure}

The temperature dependence of the "quasi-condensate" $N_{{Qbec}}$ is, see figure \ref{CondensateFract}:
\begin{equation}\label{BEC-III-sg}
\frac{N_{{Qbec}}}{{N}} =
\left\{ \begin{array}{ll}
1 - \left(\frac{T}{T_{c}}\right)^{3} \ , & \textrm{~~~~~${T}_{m} \leq T \leq {T}_{c}$ }, \\
\frac{{{\tau}}^2}{{T}_{c}^{3}}T \ \ \ \ \ \ ,  & \textrm{~~~~~$ T \leq {T}_{m}$ }.
\end{array} \right.
\end{equation}
The corresponding ground state conventional BEC is, see figure \ref{CondensateFract}:
\begin{equation}\label{BEC-I-sg}
\frac{N_{{bec}}}{{N}} =
\left\{\begin{array}{ll}
0 \ , & \textrm{${{T}}_{m} \leq T \leq {{T}}_{c}$}, \\
1 - (\frac{T}{T_c})^{3}\left(1 +\frac{{\tau}^2}{T^{2}}\right),  & \textrm{$ T \leq {{T}}_{m}$},
\end{array} \right.
\end{equation}
and for the two coexisting condensates one gets:
\begin{equation}\label{BEC-Tot-sg}
\frac{N - N_c}{N}= \frac{N_{0}}{N}= \frac{N_{Qbec}+N_{bec}}{N} = 1 - \left(\frac{T}{T_c}\right)^{3/2}.
\end{equation}

\noindent 4. Another physical observable to characterize this second critical temperature
is the correlation function condensate coherence length.
In quasi-1D harmonic trap, the correlation function is locally defined by:
{\begin{equation}\label{2-point-corr}
g(r,r')=\sum_{s} \frac{\overline{\phi}_{s}(r) \phi_{s}(r')}{e^{\beta(\epsilon_s -\mu)} - 1} \ ,
\end{equation}
where the wave function $\phi_{s}(r)$ are the well known three dimensional hermite functions:
\begin{eqnarray*}
\phi_{s}(r)=\phi_{s_x}(x)\phi_{s_y}(y)\phi_{s_z}(z),
\end{eqnarray*}
with for $\nu=x,y,z$ ($r_x\equiv x,r_y\equiv y,r_z\equiv z$):
\begin{eqnarray*}
\phi_{s_\nu}(r_\nu)&=&c_{s_\nu}\frac{1}{\sqrt{L_{\nu}}}H_{s_\nu}(r_\nu/L_\nu)e^{-\frac{r_\nu^2}{2L_\nu^2}},
\end{eqnarray*}
with $c_{s_\nu}=1/\sqrt{s_\nu! 2^{s_\nu}\sqrt{\pi}},\ \nu=x,y,z$ and
where the effective confinement lengths $L_\nu,\ \nu=x,y,z$ are given by:
$L_\nu=\sqrt{\frac{\hbar}{m\omega_\nu}}$.
Notice that the limiting diagonal function $n(r)\equiv \sigma(r,r)$ is \textit{local} space particle density at point $r$.

To compute the correlations in the direction $z$, induced by the quasi-condensate,
we have to consider $r=\widetilde{r}-\Delta r/2$ $r'=\widetilde{r}+\Delta r/2$, where $\Delta r/2=(0,0,\Delta z)$ and integrate $\widetilde{r}$
over the space to keep the effect of condensate through the trap, it give the mean value of correlation for the distance $\Delta r$:
\begin{eqnarray}\label{gmoy}
&&\langle{g}\rangle(\Delta r)=\int_{\mathbb{R}^3}d^3\widetilde{r} {g}(\widetilde{r}-\Delta r/2,\widetilde{r}+\Delta r/2).
{}\nonumber\\{}&=&\langle{g_{Nbec}}\rangle(\Delta r)+\langle{g_{Qbec}}\rangle(\Delta r)+\langle{g_{bec}}\rangle(\Delta r)
\end{eqnarray}
where correlation function is divided in three parts, each part is the truncation of the sum (\ref{2-point-corr})
: the non-condensate correlations $\langle{g_{Nbec}}\rangle\equiv\sum_{s_x>0,s_y>0,s_z}$,
the quasi-condensate correlations $\langle{g_{Qbec}}\rangle\equiv\sum_{s_x=0,s_y=0,s_z}$
and the condensate correlations $\langle{g_{bec}}\rangle\equiv$ (correlation of the ground state).
By formulas (\ref{2-point-corr}) and (\ref{gmoy})
and by mathematical assumptions \cite{A-S}, \cite{Beau2} for the non-condensate part, we get, see figure \ref{gBEC}:
\begin{eqnarray}\label{sigmaNBEC}
&&\langle{g_{Nbec}}\rangle(\Delta r)\simeq\int_{\mathbb{R}^1}d\widetilde{x}\int_{\mathbb{R}^1}d\widetilde{y}\int_{\mathbb{R}^1}d\widetilde{z}
\sum_{j=1}^{\infty}\frac{e^{j\beta\mu}}{\sqrt{\pi j\hbar\omega_0/2}^3}
{}\nonumber\\{}&\times&e^{-j m\beta(\omega_x^2\widetilde{x}^2+\omega_y^2\widetilde{y}^2+\omega_z^2\widetilde{z})^2}
e^{-2m\frac{(\Delta r)^2}{j\beta\hbar^2}}
{}\nonumber{}\\{}&=&\sum_{j=1}^{\infty}e^{j\beta\mu}\frac{1}{j^3\hbar^3\omega_0^3}\exp[-2m\frac{(\Delta r)^2}{j\beta\hbar^2}],
\end{eqnarray}
which decrease exponentially fast for $\mu<0$.
The quasi-condensate part, for $T<T_c$, by the scaling argument of localization of the scale of energetic momentum of quasi-condensate
given by $\Delta\mu$, and by mathematical assumptions \cite{A-S}, \cite{Beau2}, is given by, see figure \ref{gBEC}:
\begin{eqnarray}\label{sigmaQBEC}
&&\langle{g_{Qbec}}\rangle(\Delta r)\simeq\int_{\mathbb{R}^1}d\widetilde{z}
\sum_{j=1}^{\infty}e^{j\beta\Delta\mu}\frac{1}{\sqrt{\pi j\hbar\omega_0/2}}
{}\nonumber\\{}&\times&\exp[-j m\beta(\omega_z^2\widetilde{z})^2\beta\Delta\mu]
\exp[-2m\frac{(\Delta r)^2\beta\Delta\mu}{j\beta\hbar^2}],
{}\nonumber{}\\{}&=&\sum_{j=1}^{\infty}e^{j\beta\mu}\frac{1}{j\hbar\omega_z}\exp[-2m\frac{(\Delta z)^2\Delta\mu}{j\beta\hbar^2}].
\end{eqnarray}
The condensate part, for $T<T_m$ is given by:
\begin{eqnarray}\label{sigmaBEC}
&&\langle{g_{bec}}\rangle(\Delta r)=\int_{\mathbb{R}^3}d\widetilde{r}\phi_{000}(\Delta r)\phi_{000}(-\Delta r)N_{bec}
{}\nonumber{}\\{}&=&N_{bec}\exp{(\frac{\Delta z^2}{L_z^2})}\simeq N_{bec}.
\end{eqnarray}
\begin{figure}
\begin{center}
  \includegraphics[scale=0.15]{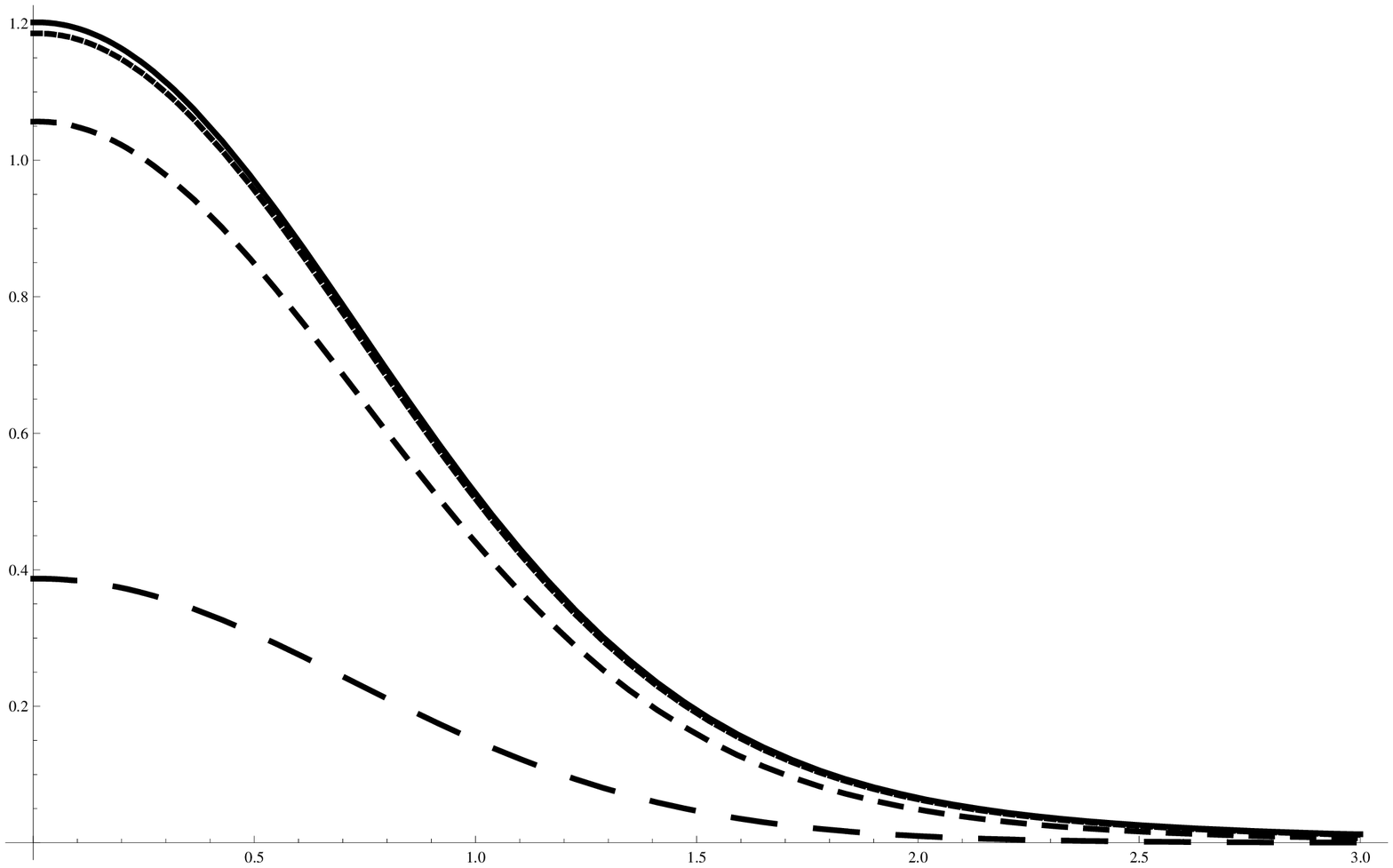}
  \includegraphics[scale=0.15]{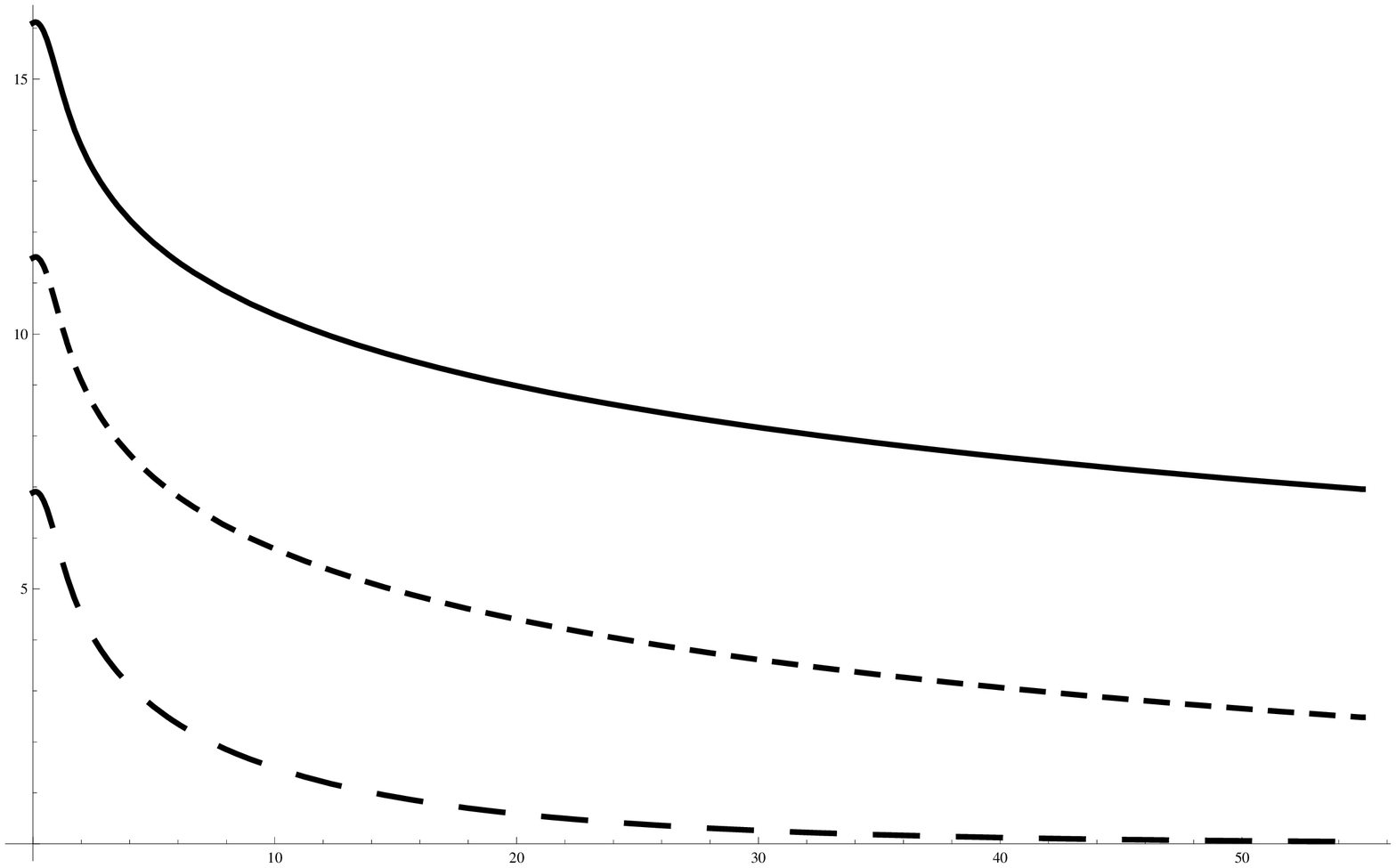}
  \caption{(a) Allure of correlation functions (\ref{sigmaQBEC}) with $\omega_0=1$, $\beta\hbar^2/2m=1$
  and $\beta\Delta\mu=1,0.1,0.01,0.001,0.00001,0.0000001$ respectively curves (-  -  -), (- - -), (...) and (---).
  Notice that (...) and (---) are combined,
  what can be interpreted as a saturation of the coherence of the thermal gas.
  (b) Allure of correlation functions (\ref{sigmaNBEC}) with $\omega_0=1$, $\beta\hbar^2/2m=1$
  for $\beta\Delta\mu=0.001,0.0001,0.000001$ respectively (-  -  -), (- - -), (---).}\label{gBEC}
\end{center}
\end{figure}
We can defined the coherence length of the quasi-condensate as the maximal scale of length of $\Delta z$
such that the correlations are non-negligible when $\omega_z, \omega_\perp$ tends to zero.
Then by a scaling argument, see equation (\ref{sigmaQBEC}) we obtain ${L}_{c}=\sqrt{\hbar/m\omega_\perp\beta\Delta\mu}$,
where $\Delta\mu$ is given by $\beta^{-1}e^{\hbar\beta\omega_z(N-N_c)}$ for $N_c<N<N_m$.
Thus $L_{c}\sqrt{m\omega_\perp/\hbar}=(\omega_\perp/\omega_z)^{\gamma(T)}$, where the exponent ${\gamma}(T)=\hbar\beta\omega_0^3(N-N_c)/\omega_c^2$
for $N_c<N<N_m$ and $\gamma(T)=\hbar\beta\omega_0^3(N_m-N_c)/\omega_c^2$ for $N>N_m$.
Using relations (\ref{Tc}) and (\ref{Tm}) between $N$, $N_c$, $N_m$ and $T$, ${T}_c$, ${T}_m$,
we can find the temperature dependence of the exponent:
\begin{eqnarray}\label{gamma}
{\gamma}(T)&=&\left(\frac{T}{\tau}\right)^2\left(\left(\frac{T_c}{T}\right)^3-1\right)\ ,\ {T}_{m}<T<{T}_{c} \ , \nonumber \\
&=&1,\ T \leq {T}_{m} \ .
\end{eqnarray}}

\begin{figure}
\begin{center}
  \includegraphics[scale=0.2]{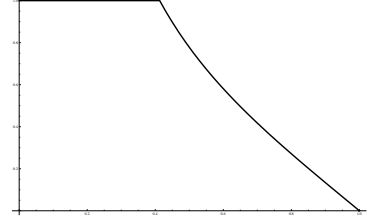}
  \caption{curve illustrating the exponent (\ref{gamma}) as a function of $\frac{T}{T_c}$ and with $\frac{T_c}{\tau}=\frac{2}{3}$.
  This gives us a relative difference between the two critical densities of 0.587, see relation \ref{TmTm} and figure \ref{C3FigTmTc}}\label{gammaH}
\end{center}
\end{figure}

{\noindent 7. We propose an experimental investigation of a weakly interacting Bose gas in quasi-one-dimensional-harmonic-trap,
with the opposite of Thomas-Fermi regime
to verify experimentally the effect of geometry on the kinetic energy and on the coherence properties of the condensate between these two critical points.
The next work will deal with the implementation of the interactions between particles and its effect on the second critical temperature
as it is done for the first critical temperature \cite{Holzmann 2}.
An other question for quasi-two-dimensional boxes is the validity of Hohenberg theorem \cite{MHL} which forbidden the existence
of the condensate on the ground state for homogeneous system.}

We are thankful to Markus Holtzmann for useful discussions about the exponential anisotropy effect.


\end{document}